\documentclass[12pt,preprint]{aastex}

\usepackage[latin1]{inputenc}
\usepackage[spanish]{babel}
\usepackage{amsmath}
\usepackage{amsfonts}
\usepackage{amssymb}
\usepackage{lscape}
\usepackage{afterpage}

\addto\captionsspanish{

    \def\abstractname{ABSTRACT}}

\usepackage[active]{srcltx}

\usepackage{fancyhdr}
\usepackage{rotating}
\usepackage{url}

\shorttitle{The ALHAMBRA photometric system}
\shortauthors{Aparicio Villegas et al.}

\begin{document}

\title{The ALHAMBRA photometric system}

\author{T. Aparicio Villegas$^{1}$, E. J. Alfaro$^1$,
  J. Cabrera-Caño$^{2,1}$, M. Moles$^{1,3}$, N. Benítez$^{1}$,
  J. Perea$^{1}$, A. del Olmo$^{1}$, A. Fernández-Soto$^{4}$,
  D. Cristóbal-Hornillos$^{1,3}$, C. Husillos$^{1}$,
  J. A. L. Aguerri$^{5}$, T. Broadhurst$^{6}$, F. J. Castander$^{7}$,
  J. Cepa$^{5,8}$, M. Cerviño$^{1}$, R. M. González Delgado$^{1}$,
  L. Infante$^{9}$, I. Márquez$^{1}$, J. Masegosa$^{1}$,
  V. J. Martínez$^{10,11}$, F. Prada$^{1}$, J. M. Quintana$^{1}$,
  S. F. Sánchez$^{3,12}$}

\affil{$^{1}$ Instituto de Astrofísica de Andalucía (CSIC) , E-18080,
  Granada, Spain; terenz@iaa.es, emilio@iaa.es, benitez@iaa.es,
  jaime@iaa.es, chony@iaa.es, cesar@iaa.es, mcs@iaa.es, rosa@iaa.es,
  isabel@iaa.es, pepa@iaa.es, fprada@iaa.es, quintana@iaa.es}

\affil{$^{2}$ Facultad de Física. Departamento de Física Atómica,
  Molecular y Nuclear. Universidad de Sevilla, Sevilla, Spain;
  jcc-famn@us.es}

\affil{$^{3}$ Centro de Estudios de Física del Cosmos de Aragón
  (CEFCA), C/ General Pizarro, 1, 44001 Teruel, Spain; moles@cefca.es,
  dch@cefca.es, sanchez@cefca.es}

\affil{$^{4}$ Instituto de Física de Cantabria (CSIC-UC), 39005,
  Santander, Spain; fsoto@ifca.unican.es}

\affil{$^{5}$ Instituto de Astrofísica de Canarias, La Laguna,
  Tenerife, Spain; jalfonso@iac.es}

\affil{$^{6}$ School of Physics and Astronomy, Tel Aviv University,
  Israel; tjb@wise.tau.ac.il}

\affil{$^{7}$ Institut de Ciències de l'Espai, IEEC-CSIC, Barcelona,
  Spain; fjc@ieec.fcr.es}

\affil{$^{8}$ Departamento de Astrofísica, Facultad de Física,
  Universidad de la Laguna, Spain; jcn@iac.es}

\affil{$^{9}$ Departamento de Astronomía, Pontificia Universidad
  Católica, Santiago, Chile; linfante@astro.puc.cl}

\affil{$^{10}$ Departament d'Astronomía i Astrofísica, Universitat de
  València, Valencia, Spain; vicent.martinez@uv.es}

\affil{$^{11}$ Observatori Astronòmic de la Universitat de València,
  Valencia, Spain}

\affil{$^{12}$ Centro Astronómico Hispano-Alemán, Almería, Spain}

\begin{abstract}
  This paper presents the characterization of the optical range of the
  ALHAMBRA photometric system, a 20 contiguous, equal-width,
  medium-band CCD system with wavelength coverage from 3500\AA{} to
  9700\AA{}. The photometric description of the system is done by
  presenting the full response curve as a product of the filters, CCD
  and atmospheric transmission curves, and using some first and second
  order moments of this response function. We also introduce the set
  of standard stars that defines the system, formed by 31 classic
  spectrophotometric standard stars which have been used in the
  calibration of other known photometric systems, and 288 stars, flux
  calibrated homogeneously, from the Next Generation Spectral Library
  (NGSL). Based on the NGSL, we determine the transformation equations
  between Sloan Digital Sky Survey (SDSS) \textit{ugriz} photometry
  and the ALHAMBRA photometric system, in order to establish some
  relations between both systems. Finally we develop and discuss a
  strategy to calculate the photometric zero points of the different
  pointings in the ALHAMBRA project.
\end{abstract}


\section{Introduction}

The ALHAMBRA (Advanced Large, Homogeneous Area Medium Band Redshift
Astronomical) survey is a project aimed at getting a photometric data
super-cube, which samples a cosmological fraction of the universe with
enough precision to draw an evolutionary track of its content and
properties (see Moles et al. 2008, for a more detailed description of
the scientific objectives of the project).  The global strategy to
achieve this aim is to identify a large number of objects, families or
structures at different redshifts (z), and compare their properties as
a function of z, once the physical dispersion of those properties at
every z-value has been taken into account. Thus, it is possible to
identify epochs (measured by z) of development of some families of
objects or structures, or their disappearance, or even the progressive
change in the proportion of a family of objects and the degree of
structuring of the universe. Such a survey requires a combination of
large area, depth, photometric accuracy and spectral coverage. The
number, width and position of the filters composing an optimal filter
set to accomplish these objectives have been discussed and evaluated
by Benítez et al. (2009), yielding a filter set formed by a uniform
system of 20 constant-width, non-overlapping, medium-band filters in
the optical range plus the three standard $JHK_s$ near infrared (NIR)
bands.

The ALHAMBRA project aims at covering a minimum of 4 square degrees in
8 discontinuous regions of the sky. This implies the need for a
wide-field instrument, associated to a 3-4m telescope that allows to
cover the selected area with the required signal-to-noise ratio,
within a reasonable observing time period. The ALHAMBRA optical system
was then incorporated to the wide-field camara
LAICA\footnote{\url{http://www.caha.es/CAHA/Instruments/LAICA/index.html}}
(Large Area Imager for Calar Alto) placed at the prime focus of the
3.5m telescope at Calar Alto Observatory, while the NIR range was
observed with the camera
OMEGA-2000\footnote{\url{http://www.caha.es/CAHA/Instruments/O2000/index.html}}
attached to the same telescope.

The main purpose of this paper is to present the characterization of
the 20 ALHAMBRA filters in the optical range, that covers from
3500\AA{} to 9700\AA{} at intervals of approximately 310\AA{}, to
select the best set of standard stars that determine the observational
system, to establish the first transformation equations between
ALHAMBRA and other photometric systems already defined, and to check
the strategies for the determination of the zero points in the
automatic reduction of the ALHAMBRA data set. Some of these issues
have been briefly presented in a previous paper introducing the
ALHAMBRA project (Moles et al. 2008).

In section 2 we present the definition of the photometric system. The
set of ALHAMBRA photometric standard stars is analyzed in detail in
section 3. In section 4.1, transformation equations between Sloan
Digital Sky Survey (SDSS) and ALHAMBRA photometric system are
established from the comparison between the synthetic magnitudes of
288 stars from the Next Generation Spectral Library (NGSL) in both
systems. We present color data for three galaxy templates as a
function of redshift using the ALHAMBRA photometric system in section 4.2, and, in section 4.3, we discuss the validity of the transformation equations for galaxies. 

Finally, in section 5 we develop a strategy to determine the
photometric zero points of the system with uncertainties below a few
hundredths of magnitude, to be applied in the calibration of the eight
ALHAMBRA fields.

\section{Characterization of the ALHAMBRA photometric system}

The ALHAMBRA photometric system has been defined according to what is
specified in Benítez et al. (2009), as a system of 20 constant-width,
non-overlapping, medium-band filters in the optical range (with
wavelength coverage from 3500\AA{} to 9700\AA{}). The definition of
the system is given by the response curve defined by the product of
three different transmission curves: the filter set, detector and
atmospheric extinction at 1.2 airmasses, based on the CAHA
monochromatic extinction tables (Sánchez et al. 2007). The graphic
representation of the optical ALHAMBRA photometric system is shown in
Fig.1. Numerical values of these response curves can be found at the
web page \url{http://alhambra.iaa.es:8080}.

First and second order moments of the transmission functions of the
different 20 filters are also of interest. We will characterize the
system filters using their isophotal wavelengths, full width half
maxima, and other derived quantities as follows.

Let $E_\lambda$ be the spectral energy distribution (SED) of an
astronomical source and $S_\lambda$ the response curve of a filter in
the photometric system defined as:
\begin{equation}
S_{\lambda} = T_t(\lambda)T_f(\lambda)T_a(\lambda),
\end{equation}
where $T_t$ is the product of the throughput of the telescope,
instrument and quantum efficiency of the detector, $T_f$ is the filter
transmission and $T_a$ is the atmospheric transmission. Following the
definition in Golay (1974), assuming that the function $E_\lambda$ is
continuous and that $S_\lambda$ does not change sign in a wavelength
interval $\lambda_a - \lambda_b$, the mean value theorem states that
there is at least one $\lambda_i$ inside the interval $\lambda_a -
\lambda_b$ such that:
\begin{equation}
  E_{\lambda_i}\int_{\lambda_a}^{\lambda_b} S_{\lambda}d\lambda = \int_{\lambda_a}^{\lambda_b} E_{\lambda}S_{\lambda}d\lambda  \label{iso}
\end{equation}
\noindent implying,
\begin{equation}
  E_{\lambda_i}=\langle E_{\lambda} \rangle = \dfrac{\int_{\lambda_a}^{\lambda_b} E_{\lambda}S_{\lambda}d\lambda}{\int_{\lambda_a}^{\lambda_b} S_{\lambda}d\lambda},  \label{iso}
\end{equation}
where $\lambda_i$ is the \textit{isophotal wavelength} and $\langle
E_{\lambda} \rangle$ denotes the mean value of the intrinsic flux
above the atmosphere over the wavelength interval (see also Tokunaga
\& Vacca 2005). Table 1 shows the isophotal wavelengths of the
ALHAMBRA filters using the spectrum of \textit{Vega} as reference. The
absolute calibrated spectrum has been taken from the STScI Observatory
Support Group
webpage\footnote{\url{http://www.stsci.edu/hst/observatory/cdbs/calspec.html}},
as the file \textit{$alpha\_lyr\_stis\_004.fits$}. This spectrum is a
combination of modeled and observed fluxes consisting of STIS CCD
fluxes from 3500\AA{}-5300\AA{}, and a Kurucz model with $T_{eff}=9400
K$ (Kurucz 2005) from 5300\AA{} to the end of the spectral range. The
isophotal wavelengths were calculated using formula~\ref{iso}.

The determination of the isophotal wavelengths of a star in a
photometric system can be complex, specially for the filters that
contain conspicuous stellar absorption lines (see Figure 1.a). For
example, filter A394M is extremely sensitive to the Balmer jump and
filter A425M encompasses the two prominent $H_\gamma$ and $H_\delta$
lines, at 4340\AA{} and 4101\AA{} respectively. At 4861\AA{},
$H_\beta$ is in filter A491M, and at 6563\AA{} $H_\alpha$ falls on
filter A646M. The Paschen series, from 8208\AA{} at filter A829M,
appears associated to the four reddest ALHAMBRA filters (A861M, A892M,
A921M, A948M).

However, this quantity depends on the spectral energy distribution of
the emitter, thus, for the same filter it will be different for each
kind of source. The isophotal wavelength can be approximated by other
central parameters that only depend on the photometric system, such as
the wavelength-weighted average or the frequency-weighted average
(written as $c\nu^{-1}$):
\begin{equation}
 \lambda_{med}=\dfrac{\int \lambda S_{\lambda} d\lambda}{\int S_{\lambda}d\lambda}
\end{equation}
\begin{equation}
\nu_{med}=\dfrac{\int \nu S_{\nu}d(\ln\nu)}{\int S_{\nu} d(\ln\nu)},
\end{equation} 
\noindent or by the effective wavelength as defined by Schneider et al.(1983):
\begin{equation}
 \lambda_{eff}=exp[\dfrac{\int d(\ln\nu) S_{\nu} \ln\lambda}{\int d(\ln\nu)S_{\nu}}]  \label{efectiva}
\end{equation}
\noindent which is halfway between the wavelength-weighted average and
the frequency-weighted average (Fukugita et al. 1996). The differences
between the isophotal wavelength concept and these three central
approaches can be seen in Golay (1974).

The root mean square (rms) fractional widths of the filters, $\sigma$,
defined as:
\begin{equation}
 \sigma=\sqrt{\dfrac{\int d(\ln\nu)S_{\nu}[\ln(\frac{\lambda}{\lambda_{eff}})]^{2}}{\int d(\ln \nu)S_{\nu}}} \label{sigma}
\end{equation}
\noindent is useful in calculating the sensitivity of the effective
wavelength to spectral slope changes:
\begin{equation}
 \delta \lambda_{eff}=\lambda_{eff} \sigma^2\delta n,
\end{equation}
\noindent where $n$ is the local power-law index of the SED
$(f_{\nu}~\nu^n)$, and the effective band width, $\delta$, is given
by:
\begin{equation}
\delta=2(2\ln2)^{1/2}\sigma \lambda_{eff}.
\end{equation}

We also calculate the flux sensitivity quantity $Q$, defined as
\begin{equation}
Q=\int d(\ln\nu)S_{\nu}, \label{Q}
\end{equation}
\noindent which allows a quick approximation to the response of the
system to a source of known flux in the following way:
\begin{equation}
N_e = AtQf_{\nu_{eff}}h^{-1},
\end{equation}
\noindent where $N_e$ is the number of photoelectrons collected with a
system of effective area A integrating for a time $t$ on a source of
flux $f_{\nu}$, and $h$ is the Planck constant. As an example, the
number of photoelectrons collected using the ALHAMBRA photometric
system, in 1 $arcsec^2$, integrating for a time of 5000 seconds on a
source of AB=23 mag$\cdot arcsec^{-2}$ in filter A491M is 1.54$\times10^4$
photoelectrons.

Table 1 shows the values of these parameters calculated for each of
the filters in the ALHAMBRA photometric system.

\section{Standard stars system}

To calibrate the ALHAMBRA photometric system we have chosen two
different sets of primary standard stars. The first one is a set of 31
classic spectrophotometric standard stars from several libraries such
as Oke \& Gunn (1983), Oke (1990), Massey \& Gronwall (1990) and Stone
(1996), together with the main standard stars adopted by the Hubble
Space Telescope (Bohlin, 2007), and the SDSS standard $BD+174708$
(Tucker et al. 2001). This set was chosen in order to anchor the
ALHAMBRA photometric system with some standards that have been used on
important photometric systems such as the SDSS, with which we have
established some transformation equations, as explained in the
following sections. These objects are stars with no detected
variability within a few millimagnitudes scale, available
high-resolution spectra ($\leq$~5 \AA{}), and flux calibrated with
errors lower than 5\%. These standard stars are spread out all over
the sky and are bright, but they do not cover all the spectral types,
being mostly white dwarfs and main sequence A-type stars.

The second set of primary standard stars is formed by 288 stars from a
new spectral library, with good spectral resolution, all of them
obtained and flux calibrated homogeneously, and covering a wide range
of spectral types, gravities and metallicities ($3440K\leq
T_{eff}\leq44500K$, 0.45 $\leq \lg g\leq$ 4.87 and $-2\leq [Fe/H]
\leq$ 0.5). This spectral library has been developed at the Space
Telescope Science Institute; it consists of 378 high signal-to-noise
stellar spectra, and is known as HST/STIS Next Generation Spectral
Library (NGSL), (Gregg et al. 2004
\footnote{\url{http://lifshitz.ucdavis.edu/~mgregg/gregg/ngsl/ngsl.html}}). The
wavelength range of these spectra covers from 2000\AA{} to
10.200\AA{}. These stars represent a potential set of
spectrophotometric standards, being valuable in the study of
atmosphere parameters using model atmosphere fluxes. Coordinates and
physical properties of these stars can be found at
\url{http://archive.stsci.edu/prepds/stisngsl/}.

These 288 stars together with the 31 classic spectrophotometric
standards define the set of primary standards of the ALHAMBRA
photometric system. AB magnitudes in the ALHAMBRA system of these stars are 
presented in two electronic tables, Table 2 and Table 3. Nine stars (AGK81D266, BD17D4708, BD28D4211,
BD75D325, FEIGE 110, FEIGE 34, G191B2B, HZ 44, P041C) are common to
both sets. Magnitudes in the ALHAMBRA photometric system for these
nine stars derived from the two different sources show differences
lower than a few millimagnitudes over the whole filter set, indicating
that both groups of standard stars are anchored to the same physical
system. However, once we have established the similitude and stability
of the photometric zero points for both set of standards we will base
the further analysis on the NGSL catalogue, which shows a higher
degree of internal homogeneity, both in wavelength and flux
calibrations.

Magnitudes in the ALHAMBRA photometric system are set on the AB
magnitude system (Oke \& Gunn 1983),
\begin{equation}
 AB_{\nu}=-2.5\,\lg f_{\nu} -48\ldotp60,
\end{equation}
where the constant comes from the expression
\begin{equation}
  48\ldotp6=-2\ldotp5 \lg F_{0},
\end{equation}
and $F_{0}=3\ldotp63 \times 10^{-20} erg/cm^2/s/Hz $ is the flux of
\textit{Vega} at $\lambda=5500$\AA{} used by those authors. The
parameter $f_{\nu}$ is the flux per unit frequency of an object in
$erg/cm^2/s/Hz$.

The AB magnitudes will be estimated as follows:
\begin{equation}
AB_{\nu} = -2\ldotp5\,\lg\, \frac{\int f_{\nu}S_{\nu}\,d(\,\lg\nu)}{\int S_{\nu}\,d(\,\lg\nu)} - 48\ldotp6
\end{equation}
where $S_{\nu}$ is the response function of the system corresponding
to the atmosphere-telescope-filter-detector combination.

Regarding the observational part of the ALHAMBRA project, the adequate
definition of the photometric system must fit both the LAICA camera
and its geometry. LAICA's focal plane is made up of four CCDs that
partly cover a total field of 44.36' x 44.36', and it needs four
filter sets, each one univocally associated to each of the CCDs. The
standard ALHAMBRA photometric system is chosen as the one at position
number 3 of the LAICA camera, and three replicas have been built to
complete the observational setting. Comparison of the three
instrumental replicas and the ALHAMBRA photometric system, has been
performed using the \textit{Vega} spectrum above. We calculated the
absolute differences between the AB magnitudes of the star in each one
of the three replicas and the standard ALHAMBRA value, detecting that
filter A366M shows the maximum absolute difference (up to 0.05
magnitudes). In the case of filter A394M there is a maximum of 0.02
magnitudes offset, filter A425M has a 0.01 magnitudes offset, and the rest of the filters
have smaller differences.

\begin{center}
 \begin{tabular}{|c|c|c|c|c|c|} 
\hline 
 Filter& CCD1 & CCD2 & CCD3 & CCD4 &  $\Delta$max \\ \hline
 A366M & 0.97 & 0.97 & 0.96 & 0.92 & 0.05 \\  
 A394M & 0.03& 0.02& 0.02 & 0.01 & 0.02 \\ 
 A425M & -0.14& -0.13& -0.13 & -0.13 & 0.01 \\ \hline \hline
\end{tabular}
\end{center}

We have developed a statistical analysis on the distribution of
differences, using synthetic AB magnitudes of the 288 standard stars
from the NGSL. Table 4 shows the result of this analysis. Once again
filter A366M presents the largest difference in magnitude, reaching up
to 0.0689 magnitudes for some stars in the sample. For the whole
sample, the median of the differences in absolute value between filter
A366M AB magnitudes in the ALHAMBRA photometric system and those
determined from the three replicas, is 0.0174 magnitudes, with a rms
(calculated through MAD, median absolute deviation) of 0.0088. In the
other filters the median of the absolute differences is less than 0.01
magnitudes, becoming smaller as we move to redder wavelengths.

\section{ALHAMBRA-SDSS transformation equations}

\subsection{Based on the Standard Stars} 

The Sloan Digital Sky Survey is a well-known astronomical project
which has obtained deep multi-color images covering more than a
quarter of the sky. Its photometric system is formed by five
non-overlapping color bands (\textit{ugriz}) that cover the complete
optical range from 3000\AA{} to 11.000\AA{} (Fukugita et al. 1996;
Smith et al. 2002). Given that the SDSS and the ALHAMBRA photometric
systems lay on the same reference standard stars, a set of
transformation equations can be determined between both systems based
on the NGSL stars.

Based on the proximity of their effective wavelengths (Table 1) , we
select the ALHAMBRA filters which better correspond to the 5 SDSS
bands as:

 \begin{center}
 \begin{tabular}{|c|l|c|l|} 
 \hline
 u& 354.0 nm& A366M & 366.1 nm \\ \hline
 g& 477.0 nm& A491M & 491.3 nm \\ \hline 
 r& 622.2 nm& A613M & 613.4 nm \\ \hline
 i& 763.2 nm& A770M & 769.9 nm \\ \hline
 z& 904.9 nm& A921M & 920.7 nm  \\ \hline
\end{tabular}
\end{center}

In this way, a zero-order approximation to the SDSS photometric system
could be given by the ALHAMBRA colors A366M-A491M (corresponding to the
u-g SDSS color), A491M-A613M (g-r color), A613M-A770M, (r-i color) and
A770M-A921M (i-z color).

However, one has to take into account the different widths of the
filters in the two systems (see Figure 1.b for a graphic
comparison). The ALHAMBRA photometric system uses medium-band filters
($\approx$ 31 nm wide), so they are more sensitive to discontinuities
in the spectral energy distribution. On the contrary, the SDSS
wide-band filters transmit more energy, permitting the detection of
fainter objects, but showing more difficulties in the interpretation and
analysis of the information derived from their photometric
measurements.

The transformation equations from SDSS magnitudes to the ALHAMBRA
photometric system (and vice versa) have been calculated using
synthetic AB magnitudes for the NGSL subset of the ALHAMBRA primary
standards. The \textit{ugriz} synthetic magnitudes were obtained from
the filters and detector set, following the SDSS standard system given
by Smith et al. (2002, USNO), located in the US Naval Observatory 1m
telescope.

The elaboration of the lineal model used to obtain the magnitudes in
the ALHAMBRA photometric system from SDSS photometry has been
developed using the statistical software {\bf R}
\footnote{\url{http://www.r-project.org}}, in particular the
statistical task called ``Backward Stepwise Regression'' that uses the
Bayesian Information Criterion (BIC), a tool for model selection
defined as:
\begin{equation}
BIC=-2\ln~L + k~\ln(n)
\end{equation}
where $L$ is the maximized value of the likelihood function for the
estimated model, $k$ the number of free parameters to be estimated,
and $n$ is the sample size (number of observations).

The transformation equations are, in general, valid for the range of
stellar parameters covered by the selected 288 NGSL stars. Table 5
shows the coefficients of the transformation from SDSS magnitudes to
ALHAMBRA photometry. First column are the dependent variables of the
linear regression which are composed of a color formed with an
ALHAMBRA band minus a SDSS band; the independent variables are
correlative SDSS colors. We calculate this type of transformations,
instead of using as dependent variables the AB ALHAMBRA magnitudes and
as the independent variables the SDSS magnitudes directly, in order to
force the sum of the linear model coefficients to be 1, a necessary
condition when we are comparing magnitudes and not colors. From a
model such as,
\begin{equation}
 Alh_i=\sum_j{a_{ij}\cdot Sloan_{j}} + b_i
\end{equation}
follows that the same star observed (using any photometric system) at
different distances would have the same magnitudes plus a single
constant equal for all filters, $\mu$, so the equations should verify
\begin{equation}
\begin{array}{rcl}
  Alh_i + \mu & = & \sum_j{a_{ij}\cdot (Sloan_j + \mu)} + b_i = \\ & = & \sum_j{a_{ij}\cdot Sloan_j} + b_i +(\sum_j{a_{ij}})\cdot \mu
\end{array}
\end{equation}
which implies that the coefficients in the transformation equations
should satisfy that the sum of all the coefficients is equal to
1. With this in mind, the equations can be posed as:
\begin{equation}
 Alh_i - Sloan_k = \sum_{j=1}^4{c_{ij}\cdot (Sloan_j-Sloan_{j+1})} + d_i
\end{equation}
where $Sloan_k$ is any of the five SDSS bands.

Table 6 shows the coefficients of the inverse transformation
equations, from ALHAMBRA photometry to SDSS magnitudes, obtained with
the same procedure.

\subsection{Redshift Galaxy templates}

Color transformations for stars, in general, would not offer accurate
results for galaxies, especially when the effects of redshift and
intergalactic absorption are taken into account. It is therefore
desirable to describe the colors of observed galaxies at different
redshifts in the system, as accurately as possible. To do so, we use
the empirically calibrated library of Benítez et al. (2004). We
generate estimated colors for the galaxies in the library in redshift
intervals of 0.05, using the SDSS filter $r$ as reference for all the
ALHAMBRA medium-band filters. All colors correspond to the AB system,
and were calculated using the functions included in the module
bpz\_tools from the BPZ package (Benítez 2000) which offers an
accuracy similar to that of SYNPHOT (Baggett et al. 1997). We present
here the tables for the E/S0, Sbc, Im, SB2, SB3 and Scd templates for the ALHAMBRA
photometric system (Tables 7 to 12). These values allow us to determine in an easy way the
$k$-correction which arises from estimating spectral indices at
observing frequencies other than the rest frequency. As an example,
Figure 2 and Figure 3 present the variation of the $k$-correction for
two ALHAMBRA colors, A491M-A613M and A770M-A921M respectively, versus
different values of z for these three galaxy templates.

These tables must be used with caution, since it is obvious that a few
templates cannot represent all the possible variations of real galaxy
spectra. Nevertheless, they also serve as a guide to transform galaxy
colors between different photometric systems for a given spectral type
and redshift. To this end, we need to find the closest possible match
in redshift and spectral type in the electronic tables, using as many
colors as possible. Then we look up the corresponding color in the
table corresponding to the spectral type and filter of interest, using
interpolation between spectral types and redshifts if needed.

\subsection{Validity of the transformation equations on galaxies}

Although the transformation equations presented in section 4.1 are
calculated to be applied to stellar objects, we have carried out a
quick test to see how they work for galaxy templates at z=0. Using
tables 7, 8 and 9, we have obtained ALHAMBRA colors of the three galaxy templates:
 E/S0, Sbc and Im. Applying the transformation equations to their ALHAMBRA colors, we obtained SDSS
colors, and these were compared with two different results presented
by Fukugita et al.(1995), where they obtained synthetic colors from
the convolution of Kennicutt's (1992) spectrophotometric atlas with
the response functions of the SDSS photometric system, and Shimasaku
et al.(2001), where SDSS colors of a sample of observed galaxies are
presented.  As Fig. 4 shows, the values obtained by the ALHAMBRA-SDSS
transformation equations are within the observed range of SDSS galaxy
colors, taking into account the typical uncertainty in the colors that
define a class of morphological type.

The validity of these transformation equations for galaxies depends on
the proximity of their spectral energy distributions to the SED of the
stars used in the elaboration of the equations, in our case, the 288
stars from the NGSL. Thus, in a color-color diagram, the smaller the
distance of the galaxy to the distribution of the stars, the higher
the confidence level will be when applying the transformation
equations to this galaxy. Figure 5 shows two ALHAMBRA color-color
diagrams where the same previous three galaxy templates are presented
at different redshift values. E/S0 and Sbc galaxies at z=0 are quite
close to the stellar distribution, while Im galaxies at z=0 are too
blue, and slightly offset from the distribution of the stellar colors.

\section{Zero point determination}

The zero point calibration of the observed fields in the ALHAMBRA
survey has been carefully considered until we found a strategy which
is robust and general enough to be applied to the whole survey and to
all kind of astronomical objects. We tackled the issue in different
ways, reaching a strategy with which zero points of ALHAMBRA fields
are determined to an error below a few hundredths of a magnitude for
the whole range of spectral types in the sample. This method entails
the following steps:
\begin{enumerate}
\item The first task is to perform an exhaustive selection of field
  stars found in the ALHAMBRA source catalogue for each field,
  selecting stellar objects with low photometric error (less than 0.15
  magnitudes), morphologically classified by SExtractor (Bertin \&
  Arnouts 1996) as stellar object at the three reddest filters, A892M,
  A921M, and A948M. It is possible that the pipeline may recognize the
  nucleus of a distant galaxy as a star in the bluest filters, so we
  focus primarily on classification in the reddest fitlers. As we show
  in Fig. 6, the great majority of these stars are main sequence stars
  of late spectral types.
\item Using this set of stars we look for the ones that have SDSS
  photometry. For this task, the SDSS DR7 was used. We utilize the
  SkyServer Tools from its web
  page\footnote{\url{http://cas.sdss.org/dr7/en/tools/search/radial.asp}}.
  Amongst the different types of search requests we choose the more
  complete SQL Search, where one can search for the needed objects
  directly from the DR7 database.
\item Then, for each stellar object, we find the star (within the 288
  NGSL objects) that best fits its SDSS photometry. To do it we select
  the NGSL star that minimizes the variance of the differences between
  the SDSS (model) magnitudes of the NGSL star and the SDSS (observed) magnitudes of the field star.
  We only allow for a constant term, equal for every magnitude, related to the distance
  module of both stars ($\delta v = 5\log (\dfrac{r_2}{r_1}))$. Once
  we have the NGSL objects that best fit with each field star, we
  assign to it the expected ALHAMBRA AB magnitudes that correspond,
  using the already known system responses. In Fig. 7 we show, as an
  example, the spectrum of the star HD160346 in AB magnitudes,
  together with the instrumental magnitudes plus the zero points of a
  field star fitted with this NGSL star in the algorithm, and also its
  SDSS photometry.
\item Finally, the data are refined from outliers with a Chebyshev
  filter:
 \begin{equation}
   |x_{i} - \bar{x}| < k\ast \sigma
\end{equation}
where $\bar{x}$ is the central value of the differences, $k$ is a
constant, in particular we take $k=3$, and $\sigma$ is the MAD of the
data. The zero point is determined as the central value of the
distribution of differences between instrumental magnitudes and
synthetic AB magnitudes.
\end{enumerate}

Table 13 shows the main characteristics of the photometric zero points
of the ALHAMBRA field 8, pointing 1, as an example of the results
obtained when applying this strategy.

In principle, the transformation equations between SDSS and ALHAMBRA
photometry and the subsequent comparison between the instrumental
values and the ALHAMBRA standard values for the stars with photometry
in both systems, should yield the same results. However, we noticed
some systematic structure in the distribution of differences (in the
sense of instrumental minus transformed AB magnitude), which depends
on the magnitude of the stars. The differences of the brighter stars
are higher than the weaker ones, up to one magnitude ahead in some
filters. In Fig. 8 we plot the magnitude differences for filter A394M
versus the SDSS magnitude u, for the same sample of field stars
(ALHAMBRA field 8, pointing 1) used with the method presented above.
We can see that the zero point estimated with the comparison of
spectra and SEDs (see Table 13) is a good fit to the point distribution
in general, although also some of the stars with brighter u magnitude
move systematically away from the central value. This anomalous result
appears also when other photometric systems (e.g. UBVRI) are compared
to the SDSS system, and these transformations are then used to
determine the zero point in the UBVRI system from data of the SDSS
survey (as shown in Chonis \& Gaskell 2008). Those authors proposed
some possible explanations to account for the observed systematic
difference. We consider that a detailed discussion on the precision
and accuracy of the brighter star photometry in the SDSS catalogue is
out of the scope of this paper. Hence we only notice that our chosen
methodology overcomes the troubles derived from a direct comparison
between instrumental and standard ALHAMBRA values, obtained from
SDSS-to-ALHAMBRA transformations.

\section{Conclusions and Summary}

The ALHAMBRA survey is a new extragalactic survey developed with a clear
scientific objective: looking for the optimization of some variables
to perform a kind of \textit{cosmic tomography} of a portion of the
universe. It is based on a new photometric system, consisting of 20
filters covering all the optical spectral range, and the three
classical $JHK_s$ near infrared bands. The survey camera for the
optical range is LAICA, installed on the prime focus of the 3.5m
telescope of the Calar Alto Astronomical Observatory. The geometry of
this instrument, with four different CCDs, is important to understand
the ALHAMBRA observational strategy.

We present the characterization of the ALHAMBRA photometric system as
defined by the product of three different response functions:
detector, filters and atmosphere.

The set of primary standard stars which defines the ALHAMBRA
photometric system is formed by 31 classic spectrophotometric standard
stars from several libraries together with 288 stars from the Next
Generation Spectral Library, which cover a wide range of spectral
types and metallicities. Since the ALHAMBRA photometric system
observes with four sets of filter+detector combinations (associated to
each of the LAICA detectors), we have chosen one of the sets as the
definition of the system, and developed a statistical analysis of the
differences between the AB magnitudes in each one of the others. This
analysis shows that the maximum difference between magnitudes, takes
place in the bluest filter, A366M, with a median difference of
0.0174 magnitudes, being much smaller in the others 19 filters.

Transformation equations from SDSS photometry to the ALHAMBRA
photometric system (and vice versa) have been elaborated using
synthetic magnitudes of the 288 standard stars from the NGSL, making
use of ``Backward Stepwise Regression'', and the ``Bayesian Information
Criterion'' as tool for model selection. We have also shown some
examples of galaxy colors and their redshift evolution in our
system. In particular, we show the ALHAMBRA colors for three galaxy
templates (E/S0, Sbc and Im) at different redshifts, from z=0.00 to
z=2.50, with a step of 0.05 in z.

The strategy of the observational determination of the photometric
zero point has been worked out in detail, with special attention to
the problems that appear when these zero points are calculated based
on the SDSS transformation equations. These problems are also present
on other photometric systems different to our own (e.g. UBVRI). The
method shown in this paper overcomes these trouble, allowing us to
determine the zero point of any ALHAMBRA field with an error below few
hundredths of magnitude.

\begin{acknowledgements}
We thank Jesús Maíz Apellaniz for providing data and advice. 

This publication makes use of data from the SDSS.Funding for the SDSS and SDSS-II has been provided by the Alfred P. Sloan Foundation, the Participating Institutions, the National Science Foundation, the U.S. Department of Energy, the National Aeronautics and Space Administration, the Japanese Monbukagakusho, the Max Planck Society, and the Higher Education Funding Council for England. The SDSS Web Site is http://www.sdss.org/.
The SDSS is managed by the Astrophysical Research Consortium for the Participating Institutions. The Participating Institutions are the American Museum of Natural History, Astrophysical Institute Potsdam, University of Basel, University of Cambridge, Case Western Reserve University, University of Chicago, Drexel University, Fermilab, the Institute for Advanced Study, the Japan Participation Group, Johns Hopkins University, the Joint Institute for Nuclear Astrophysics, the Kavli Institute for Particle Astrophysics and Cosmology, the Korean Scientist Group, the Chinese Academy of Sciences (LAMOST), Los Alamos National Laboratory, the Max-Planck-Institute for Astronomy (MPIA), the Max-Planck-Institute for Astrophysics (MPA), New Mexico State University, Ohio State University, University of Pittsburgh, University of Portsmouth, Princeton University, the United States Naval Observatory, and the University of Washington.

Some/all of the data presented in this paper were obtained from the Multimission Archive at the Space Telescope Science Institute (MAST). STScI is operated by the Association of Universities for Research in Astronomy, Inc., under NASA contract NAS5-26555. Support for MAST for non-HST data is provided by the NASA Office of Space Science via grant NAG5-7584 and by other grants and contracts.

We acknowledge support from the Spanish Ministerio de Educación y Ciencia through grant AYA2006-14056  BES-2007-14764. E. J. Alfaro acknowledges the financial support from the Spanish MICINN under the Consolider-Ingenio 2010 Program grant CSD2006-00070: First Science with the GTC.

\end{acknowledgements}

\begin{center}
 \begin{figure}[H!]
 \centering
 \includegraphics[width=\columnwidth]{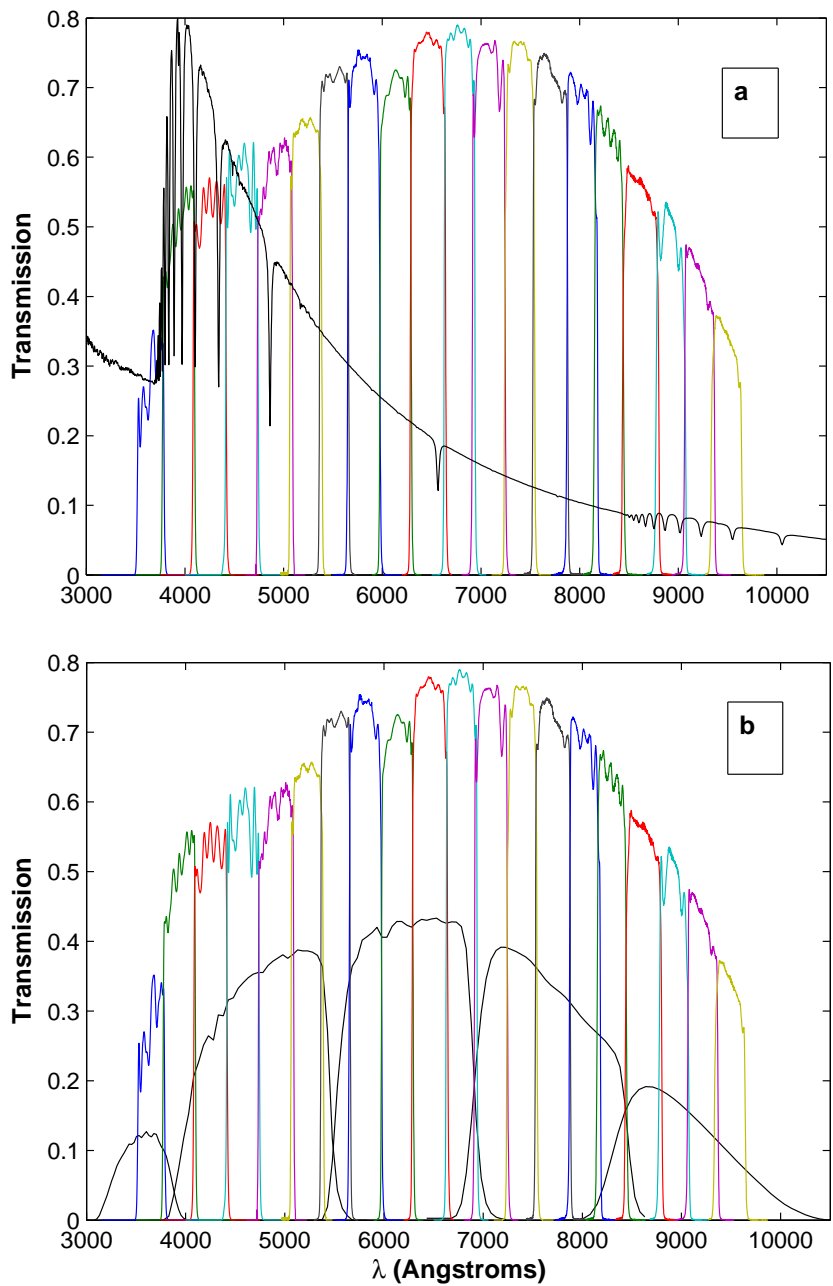}
 \caption{Response functions of the ALHAMBRA photometric system
   filters including atmospheric transmission at 1.2 airmasses at the
   altitude of Calar Alto Observatory. Figure 1.a also represents the spectrum of \textit{Vega} superimposed on the transmission curves. The flux of \textit{Vega} has been normalized and scaled
   properly to make the graphic possible. Figure 1.b shows the response curves of the SDSS standard system (USNO) and the ALHAMBRA
   photometric system.}
\end{figure}
\end{center}

 \begin{center}
 \begin{table}[H!]
  \title{
\begin{center} Table 1 \end{center} 
 \begin{center}
 Representative parameters of the ALHAMBRA photometric system filters
\end{center}
 }
 \begin{center}
\begin{footnotesize}
 \begin{tabular}{ccccccccccc}
 \hline
 & $\lambda_{iso}$ \footnotemark[1]& $F_\lambda$ & \textit{Vega} AB &$\lambda_{m}$ \footnotemark[2]& $c\nu_{m}^{-1}$ \footnotemark[3]& $\lambda_{eff}$ \footnotemark[4]& & & FWHM &  \\ 
 Filter & (nm) & $(erg/s/cm^2/$\AA{}) & Magnitude & (nm) & (nm) & (nm) & $\sigma$ & Q & (nm) & $\delta$\\ \hline
 A366M& 373.8&3.354e-09&0.96 &366.4& 366.0& 366.1& 0.0216& 0.0211 & 27.9 & 186.2\\ 
 A394M& 398.5&6.851e-09&0.02  &394.4& 393.9& 394.1& 0.0239& 0.0414 & 33.0 & 222.2\\ 
 A425M& 430.9&6.782e-09&-0.13 &425.3& 424.9& 424.9& 0.0229& 0.0419 & 34.2 & 229.8\\ 
 A457M& 456.8&6.117e-09&-0.18 &457.8& 457.4& 457.5& 0.0209& 0.0412 & 33.2 & 224.9\\ 
 A491M& 500.8&4.726e-09&-0.05 & 491.7& 491.2& 491.3& 0.0208& 0.0426 & 35.6 & 241.2\\ 
 A522M& 522.4&4.145e-09&-0.04 & 522.6& 522.3& 522.4& 0.0178& 0.0395 & 32.6 & 218.9\\ 
 A551M& 551.0&3.547e-09&0.01  & 551.2& 550.9& 551.0& 0.0156& 0.0385 & 29.7 & 202.1\\ 
 A581M& 581.1&3.031e-09&0.07  & 581.2& 580.9& 580.9& 0.0161& 0.0407 & 32.4 & 221.1\\ 
 A613M& 613.4&2.573e-09&0.13  & 613.6& 613.4& 613.4& 0.0149& 0.0364 & 32.0 & 216.2\\ 
 A646M& 650.8&2.119e-09&0.23 & 646.4& 646.0& 646.1& 0.0159& 0.0419 & 35.7 & 241.5\\ 
 A678M& 678.1&1.896e-09&0.24&  678.2& 678.0& 678.1& 0.0133& 0.0355 & 31.4 & 211.9\\ 
 A708M& 707.9&1.661e-09&0.29&  707.9& 707.7& 707.8& 0.0135& 0.0347 & 33.2 & 225.3\\ 
 A739M& 739.1&1.453e-09&0.34&  739.3& 739.1& 739.2& 0.0117& 0.0302 & 30.4 & 203.3\\ 
 A770M& 770.1&1.277e-09&0.39&  770.2& 769.9& 769.9& 0.0132& 0.0329 & 35.4 & 239.2\\ 
 A802M& 802.4&1.123e-09&0.44&  802.2& 801.9& 802.0& 0.0111& 0.0263 & 31.2 & 210.5\\ 
 A829M& 830.2&1.012e-09&0.48&  829.5& 829.3& 829.4& 0.0103& 0.0226 & 29.6 & 202.1\\ 
 A861M& 861.5&8.943e-10&0.54&  861.6& 861.3& 861.4& 0.0121& 0.0233 & 36.9 & 246.4\\ 
 A892M& 884.3&8.703e-10&0.50&  891.9& 891.7& 891.8& 0.0095& 0.0162 & 30.3 & 200.2\\ 
 A921M& 936.2&8.197e-10&0.48&  920.9& 920.7& 920.8& 0.0096& 0.0143 & 30.8 & 207.8\\ 
 A948M& 950.4&7.502e-10&0.52&  948.3& 948.2& 948.2& 0.0093& 0.0109 & 31.9 & 208.1\\ \hline \hline
\multicolumn{11}{l}{\scriptsize{1. Isophotal wavelength defined in equation \eqref{iso}}} \\
\multicolumn{11}{l}{\scriptsize{2. Wavelength-weighted average}} \\
\multicolumn{11}{l}{\scriptsize{3. Inverse of frequency-weighted average}} \\
\multicolumn{11}{l}{\scriptsize{4. Effective wavelength defined in equation \eqref{efectiva}}} \\ \hline
 \end{tabular}
 \end{footnotesize}
\tablecomments{Columns 2 to 4 represent the isophotal wavelengths, flux densities and AB magnitudes 
  of \textit{Vega} in the ALHAMBRA photometric system.}
\end{center}
\end{table}
\end{center}

\begin{deluxetable}{lccccc}
\tablecaption{ALHAMBRA AB magnitudes of the 31 spectrophotometric primary standard stars}
\tablenum{2}
\tablehead{
\colhead{Star} &
\colhead{A366M} &
\colhead{A491M} &
\colhead{A613M} &  
\colhead{A770M} &
\colhead{A921M}
}
\startdata
g191b2b\_mod\_004 & 11.04&11.57&11.99&12.45&12.82 \\
bd\_75d325\_stis\_001 & 8.77 & 9.32 & 9.75 & 10.21 & 10.58 \\
feige34\_stis\_001 & 10.40 & 10.97 & 11.38 & 11.80 & 12.11 \\
p041c\_stis\_001 & 13.40 & 12.13 & 11.84 & 11.73 & 11.72 \\
alpha\_lyr\_stis\_004 & 0.96 & -0.05 & 0.13 & 0.39 & 0.48 \\
\enddata
\tablecomments{Table 2 is published in its entirety in the electronic 
edition of the {\it Astronomical Journal}.  A small portion of columns
and rows are shown here for guidance regarding its form and content.}
\end{deluxetable}

\begin{deluxetable}{lccccc}
\tablecaption{ALHAMBRA AB magnitudes of the 288 primary standard stars from the NGSL}
\tablenum{3}
\tablehead{
\colhead{Star} &
\colhead{A366M} &
\colhead{A491M} &
\colhead{A613M} &  
\colhead{A770M} &
\colhead{A921M}
}
\startdata
BD17d4708 & 10.45 & 9.57 & 9.34 &9.25 & 9.24 \\
Feige110 & 11.15 & 11.62 & 12.05 & 12.47 & 12.82 \\
G188-22 & 11.11 & 10.24 & 9.98 & 9.83 & 9.87 \\
HD086986 & 9.13 & 7.94 & 8.02 & 8.14 & 8.16 \\
HD204543 & 10.23 & 8.56 & 7.99 & 7.63 & 7.52 \\
\enddata
\tablecomments{Table 3 is published in its entirety in the electronic 
edition of the {\it Astronomical Journal}.  A small portion of columns
and rows are shown here for guidance regarding its form and content.}
\end{deluxetable}

\begin{center}
\begin{table}[H!]
\title{\begin{center} Table 4 \end{center} 
\begin{center}
Analysis of the four instances of the ALHAMBRA system used with the LAICA camera
\end{center}
}
\begin{center}
 \begin{tabular}{cccc}
\hline
Filter & $\Delta$Max & Median & RMS \\ \hline
A366M & 0.0689 & 0.0174 &  0.0088\\ 
A394M & 0.0236 & 0.0064 &  0.0053\\ 
A425M & 0.0592 & 0.0058 &  0.0073\\ 
A457M & 0.0148 & 0.0028 &  0.0025\\ 
A491M & 0.0133 & 0.0011 &  0.0013\\ 
A522M & 0.0325 & 0.0022 &  0.0021\\ 
A551M & 0.0092 & 0.0008 &  0.0009\\ 
A581M & 0.0110 & 0.0024 &  0.0019\\ 
A613M & 0.0052 & 0.0003 &  0.0003\\ 
A646M & 0.0085 & 0.0012 &  0.0007\\
A678M & 0.0247 & 0.0009 &  0.0007\\ 
A708M & 0.0031 & 0.0002 &  0.0001\\ 
A739M & 0.0228 & 0.0009 &  0.0006\\ 
A770M & 0.0042 & 0.0003 &  0.0003\\ 
A802M & 0.0134 & 0.0006 &  0.0006\\ 
A829M & 0.0016 & 0.0003 &  0.0003\\ 
A861M & 0.0133 & 0.0009 &  0.0007\\ 
A892M & 0.0034 & 0.0002 &  0.0003\\ 
A921M & 0.0013 & 0.0003 &  0.0001\\ 
A948M & 0.0028 & 0.0005 &  0.0003\\ \hline \hline
\end{tabular}
\end{center}
\tablecomments{First column is the list of ALHAMBRA filters;
    second is the maximum absolute difference between ALHAMBRA AB
    magnitudes and the magnitudes obtained from the three
    filter-detector replicas at each ALHAMBRA filter; third represents
    the median of the former absolute differences; fourth is the
    median absolute deviation of the distribution.}
\end{table}
\end{center}

\begin{center}
\begin{table}[H!]
\title{\begin{center} Table 5 \end{center}
\begin{center}
Coefficients of the transformation equations from SDSS to ALHAMBRA photometry
\end{center}}
\begin{center}
  \begin{tabular}{ccccccc}
 \hline \hline
\multicolumn{6}{c}{Synthetic} &   \\ \cline{1-6}
 &$\emptyset$ & u-g& g-r& r-i& i-z& Error  \\ \cline{2-7}
A366M - i &-0.0209&  0.9813 & 0.7946 & 1.0889 & & 0.038\\ 
A394M - i &-0.1997&   0.4183 & 1.9673 &  0.8166 & -0.8552 & 0.097 \\ 
A425M - i &-0.0639&  0.1475 & 1.4885 & 0.8921 & & 0.058\\ 
A457M - i &-0.0094& & 1.0195 & 1.0358 & 0.1799 & 0.037\\ 
A491M- i &0.0240 &  & 0.6274 & 1.2114 & 0.1969 & 0.024\\
A522M - i &-0.0239& & 0.6204 & 1.0030 & & 0.028\\ 
A551M - g &-0.0093& -0.0108 & -0.6518 & 0.1231 & -0.0975 & 0.010\\ 
A581M - g &-0.0029&  -0.0018 & -0.9166 & 0.2327 & -0.0547 & 0.007\\
A613M - g &-0.0009& 0.0129 & -1.0759 & 0.1896 & & 0.009\\ 
A646M - g &-0.0023& 0.0159 & -1.0349 & -0.2544 &  & 0.011\\
A678M - g &0.0131& 0.0077 & -1.2133 & -0.1184 & 0.1035 & 0.022\\
A708M - g &-0.0155& 0.0218 &  -1.1169 & -0.4915 & & 0.014\\ 
A739M - g &-0.0096& -0.0104 & -0.9076 & -1.1087 & 0.0677 & 0.006\\
A770M - g &0.0165& -0.0068 & -1.0636 & -0.9769 & & 0.007\\ 
A802M - g &-0.0005& 0.0139 & -1.0063 & -1.2263 & -0.1022 & 0.006\\ 
A829M - g &0.0036& 0.0199 & -1.0274 & -1.2496 & -0.2488 & 0.006\\ 
A861M - g &0.0159& 0.0211 & -1.0723 & -1.0679 & -0.6102 & 0.011\\ 
A892M - g &0.0171& -0.0192 & -1.0259 & -1.0268 & -0.8638 & 0.015\\ 
A921M - g &0.0055& -0.0163 & -0.9414 & -1.0725 & -1.1318 & 0.019\\ 
A948M - g &0.0221& & -0.9802 & -0.9667 & -1.4006 & 0.021\\ \hline \hline
\end{tabular}
 \end{center}
\tablecomments{\hspace{0.05cm}Last column represents the residual standard error of each fit.}
  \end{table}
 \end{center}

\begin{center}
\begin{table}[H!]
\title{\begin{center} Table 6 \end{center}
\begin{center}
Coefficients of the transformation equations from ALHAMBRA to SDSS photometry
\end{center}}
\begin{center}
  \begin{tabular}{cccccc}
 \hline \hline
\multicolumn{6}{c}{Synthetic}    \\ \hline
& u - A522M & g - A613M & r - A522M & i - A522M & z - A613M \\ \cline{2-6}
$\emptyset$ & -0.0247 & 0.0197 &  -0.0006 & -0.0002 & -0.0116 \\ 
A366M - A394M & 1.0741 & & & & \\ 
A394M - A425M & 1.2574 & 0.0467 & & -0.0142 & -0.0337 \\ 
A425M - A457M & 0.9768 & 0.2250 & & 0.0177 &  \\ 
A457M - A491M & 1.4529 & 0.2282 & -0.0535 & -0.0336 & \\ 
A491M - A522M & 0.8850 & 0.4580 & & & \\ 
A522M - A551M &  & 0.6206 & -1.0334 & -1.0117 & \\ 
A551M - A581M & & 1.4129 & -0.8928 & -0.9767 &  \\ 
A581M - A613M &  & 1.3538 & -0.5793 & -0.9498 & \\ 
A613M - A646M & & & -0.3299 & -0.9424 & -0.9459 \\ 
A646M - A678M & -0.2832 & & -0.1895 & -0.9525 & -0.9399 \\ 
A678M - A708M & 1.9317 & & & -1.0060 & -1.0696 \\ 
A708M - A739M & -1.6333 & -0.4194 & -0.2017 & -0.8624 & -1.0935 \\ 
A739M - A770M & -1.7633 & & -0.1224 & -0.5364 & -0.7336 \\ 
A770M - A802M & & 0.6199 & & -0.2679 & -1.0201 \\ 
A802M - A829M & 2.5577 & 1.2026 & 0.3267 & & -0.5209 \\ 
A829M - A861M & & -0.2826 & & & -1.0507 \\ 
A861M - A892M & & & & & -0.5793 \\ 
A892M - A921M & -1.2739 & -0.2896 & -0.0429 & & \\ 
A921M - A948M & & & & & -0.7648 \\ 
Error & 0.0253 & 0.0121 & 0.0035 & 0.0024 &  0.0102 \\ \hline \hline
\end{tabular}
 \end{center}
\tablecomments{For each SDSS filter minus one of the
    ALHAMBRA filters, the column below shows the coefficients of each
    one of the 19 ALHAMBRA colors, plus the independent term. Last row
    gives the residual standard error of each fit.}
  \end{table}
 \end{center}

\begin{deluxetable}{lccccc}
\tablecaption{Multi-instrument colors of the E/S0 galaxy template from Benitez et al. (2004) in the ALHAMBRA system (AB mag) and SDSS r (AB mag) }
\tablenum{7}
\tablehead{
\colhead{z} &
\colhead{A366M-r} &
\colhead{A491M-r} &
\colhead{A613M-r} &  
\colhead{A770M-r} &
\colhead{A921M-r}
}
\startdata
0.00 & 2.26 & 0.55 & -0.01 & -0.41 & -0.66 \\
0.25 & 3.00 & 1.32 & -0.01 & -0.58 & -0.94 \\
1.00 & 3.47 & 2.26 & 0.12 & -0.88 & -2.38 \\
1.50 & 1.69 & 1.06 & 0.21 & -1.98 & -3.01 \\
2.25 & 1.17 & 0.47 & 0.09 & -0.83 & -2.48 \\
\enddata
\tablecomments{Table 7 is published in its entirety in the electronic 
edition of the {\it Astronomical Journal}.  A small portion of columns
and rows are shown here for guidance regarding its form and content.}
\end{deluxetable}

\begin{deluxetable}{lccccc}
\tablecaption{Multi-instrument colors of the Sbc galaxy template from Benitez et al. (2004) in the ALHAMBRA system (AB mag) and SDSS r (AB mag) }
\tablenum{8}
\tablehead{
\colhead{z} &
\colhead{A366M-r} &
\colhead{A491M-r} &
\colhead{A613M-r} &  
\colhead{A770M-r} &
\colhead{A921M-r}
}
\startdata
0.00 & 1.64 & 0.45 & 0.01 & -0.33 & -0.55 \\
0.25 & 2.28 & 0.76 & 0.00 & -0.44 & -0.73 \\
1.00 & 1.03 & 0.56 & 0.07 & -1.17 & -1.94 \\
1.50 & 1.22 & 0.34 & 0.01 & -0.49 & -1.46 \\
2.25 & 1.92 & 0.69 & -0.01 & -0.34 & -0.67 \\
\enddata
\tablecomments{Table 8 is published in its entirety in the electronic 
edition of the {\it Astronomical Journal}.  A small portion of columns
and rows are shown here for guidance regarding its form and content.}
\end{deluxetable}

\begin{deluxetable}{lccccc}
\tablecaption{Multi-instrument colors of the Im galaxy template from Benitez et al. (2004) in the ALHAMBRA system (AB mag) and SDSS r (AB mag) }
\tablenum{9}
\tablehead{
\colhead{z} &
\colhead{A366M-r} &
\colhead{A491M-r} &
\colhead{A613M-r} &  
\colhead{A770M-r} &
\colhead{A921M-r}
}
\startdata
0.00 & 0.75 & -0.00 & -0.00 & -0.19 & -0.30 \\
0.25 & 1.20 & 0.43 & -0.17 & -0.13 & -0.28 \\
1.00 & 0.27 & 0.15 & 0.02 & -0.63 & -1.00 \\
1.50 & 0.42 & 0.07 & 0.00 & -0.13 & -0.76 \\
2.25 & 0.95 & 0.27 & -0.01 & -0.09 & -0.16 \\
\enddata
\tablecomments{Table 9 is published in its entirety in the electronic 
edition of the {\it Astronomical Journal}.  A small portion of columns
and rows are shown here for guidance regarding its form and content.}
\end{deluxetable}

\begin{deluxetable}{lccccc}
\tablecaption{Multi-instrument colors of the SB2 galaxy template from Benitez et al. (2004) in the ALHAMBRA system (AB mag) and SDSS r (AB mag) }
\tablenum{10}
\tablehead{
\colhead{z} &
\colhead{A366M-r} &
\colhead{A491M-r} &
\colhead{A613M-r} &  
\colhead{A770M-r} &
\colhead{A921M-r}
}
\startdata
0.00 & 0.54 & -0.34 & 0.24 & -0.02 & -0.25 \\
0.25 & 1.00 & 0.44 & -0.51 & 0.21 & 0.01 \\
1.00 & 0.13 & 0.08 & 0.08 & -0.45 & -0.62 \\
1.50 & 0.06 & 0.12 & -0.06 & -0.02 & -0.71 \\
2.25 & 0.56 & 0.01 & -0.00 & -0.03 & -0.06 \\
\enddata
\tablecomments{Table 10 is published in its entirety in the electronic 
edition of the {\it Astronomical Journal}.  A small portion of columns
and rows are shown here for guidance regarding its form and content.}
\end{deluxetable}

\begin{deluxetable}{lccccc}
\tablecaption{Multi-instrument colors of the SB3 galaxy template from Benitez et al. (2004) in the ALHAMBRA system (AB mag) and SDSS r (AB mag) }
\tablenum{11}
\tablehead{
\colhead{z} &
\colhead{A366M-r} &
\colhead{A491M-r} &
\colhead{A613M-r} &  
\colhead{A770M-r} &
\colhead{A921M-r}
}
\startdata
0.00 & 1.15 & 0.32 & 0.09 & -0.11 & -0.45 \\
0.25 & 1.05 & 0.49 & -0.04 & -0.27 & -0.40 \\
1.00 & 0.36 & 0.21 & -0.02 & -0.42 & -0.96 \\
1.50 & 0.31 & 0.20 & -0.02 & -0.24 & -0.50 \\
2.25 & 0.74 & 0.22 & -0.00 & -0.11 & -0.26 \\
\enddata
\tablecomments{Table 11 is published in its entirety in the electronic 
edition of the {\it Astronomical Journal}.  A small portion of columns
and rows are shown here for guidance regarding its form and content.}
\end{deluxetable}

\begin{deluxetable}{lccccc}
\tablecaption{Multi-instrument colors of the Scd galaxy template from Benitez et al. (2004) in the ALHAMBRA system (AB mag) and SDSS r (AB mag) }
\tablenum{12}
\tablehead{
\colhead{z} &
\colhead{A366M-r} &
\colhead{A491M-r} &
\colhead{A613M-r} &  
\colhead{A770M-r} &
\colhead{A921M-r}
}
\startdata
0.00 & 1.31 & 0.30 & 0.00 & -0.22 & -0.27 \\
0.25 & 1.60 & 0.62 & -0.05 & -0.34 & -0.53 \\
1.00 & 0.61 & 0.40 & 0.01 & -0.73 & -1.35 \\
1.50 & 0.74 & 0.13 & 0.01 & -0.39 & -0.94 \\
2.25 & 1.41 & 0.48 & -0.01 & -0.15 & -0.38 \\
\enddata
\tablecomments{Table 12 is published in its entirety in the electronic 
edition of the {\it Astronomical Journal}.  A small portion of columns
and rows are shown here for guidance regarding its form and content.}
\end{deluxetable}

\clearpage

\begin{center}
 \begin{figure}[H]
  \centering
  \includegraphics[width=\columnwidth]{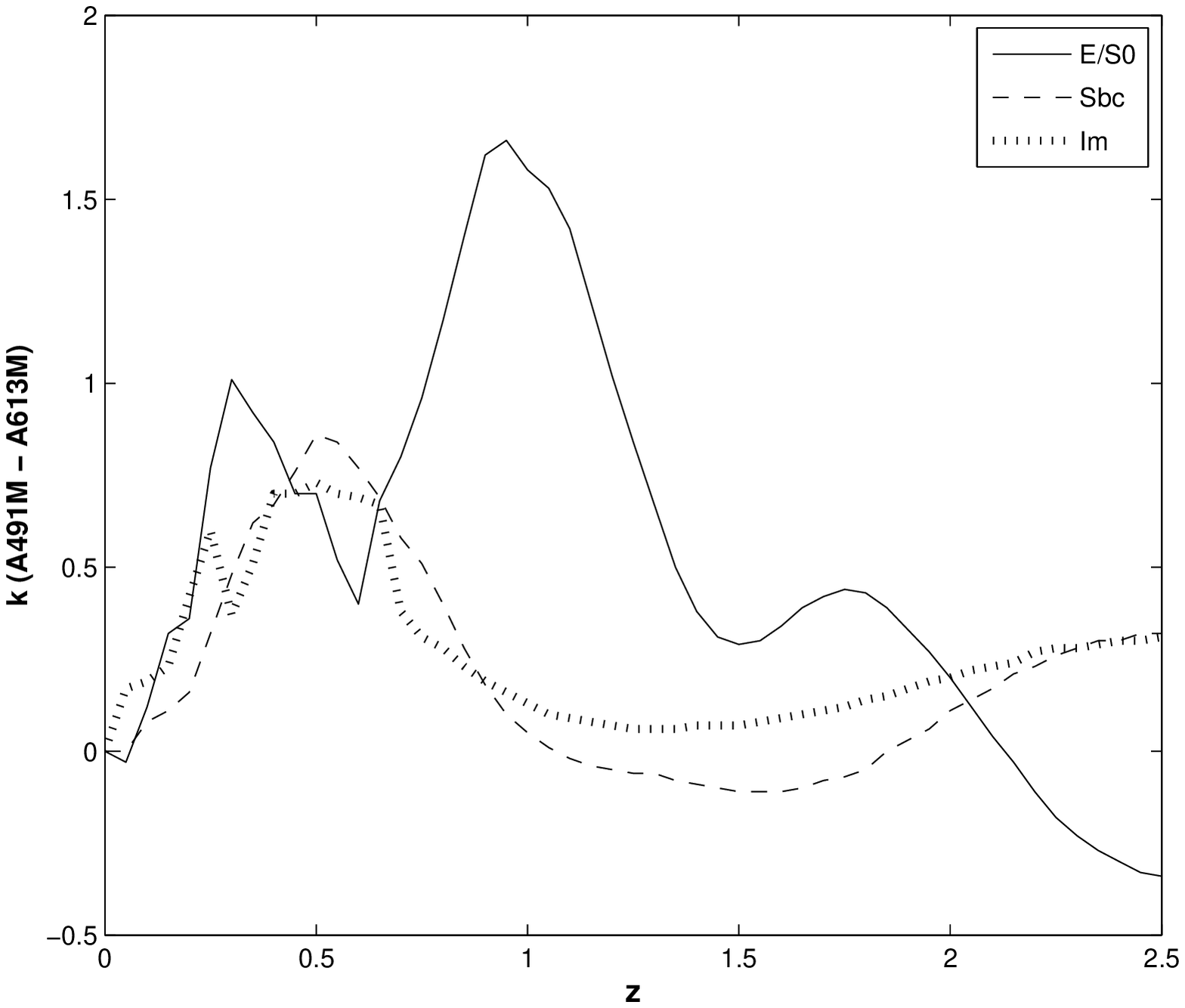}
  \caption{\footnotesize k-correction for different values of redshift
    (z) for ALHAMBRA color A491M-A613M and three galaxy templates.}
  \end{figure}
 \end{center}

\begin{center}
 \begin{figure}[H]
  \centering
  \includegraphics[width=\columnwidth]{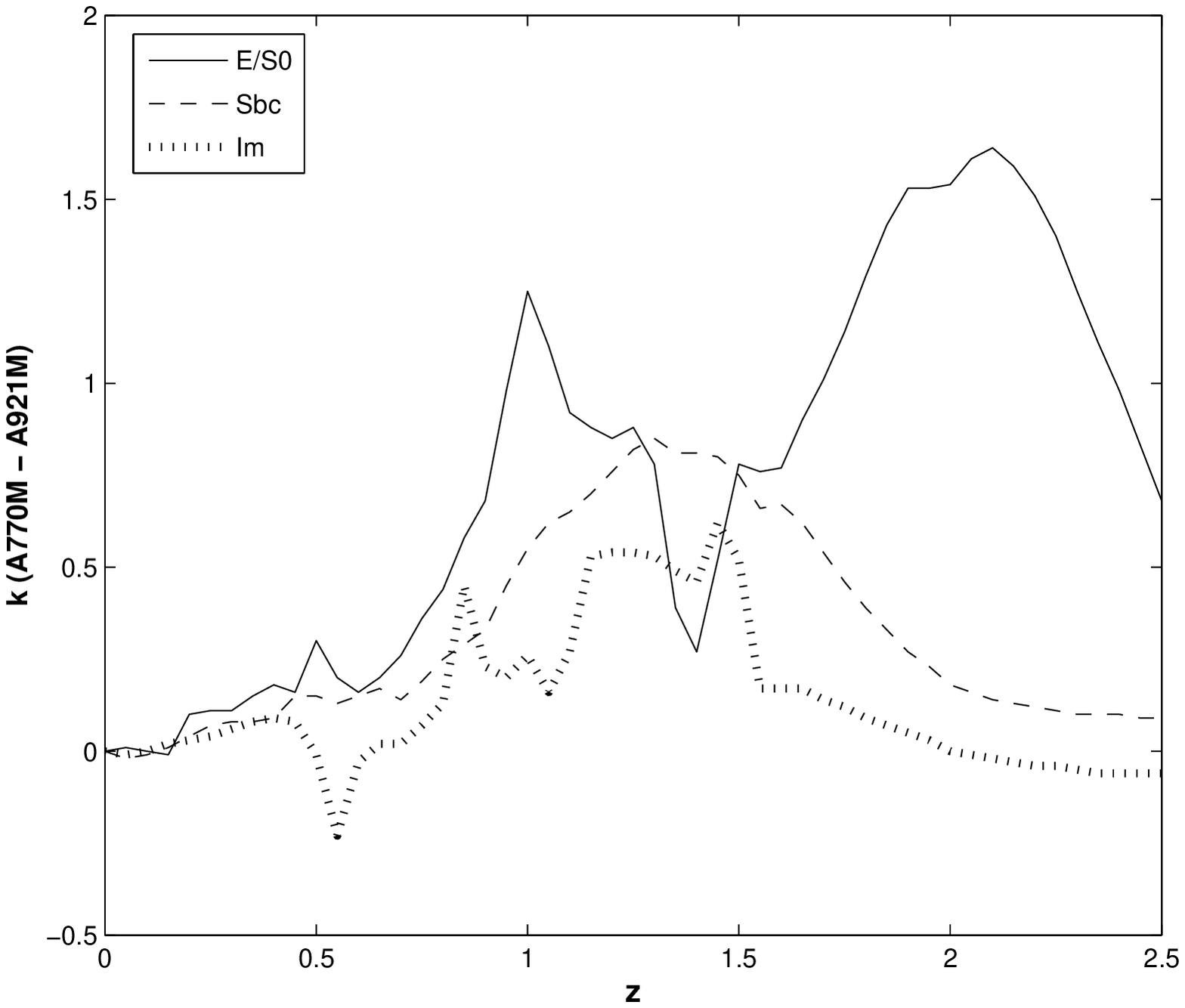}
  \caption{\footnotesize k-correction for different values of redshift
    (z) for ALHAMBRA color A770M-A921M and three galaxy templates.}
  \end{figure}
 \end{center}

\begin{center}
 \begin{figure}[H]
 \centering
 \includegraphics[width=\columnwidth]{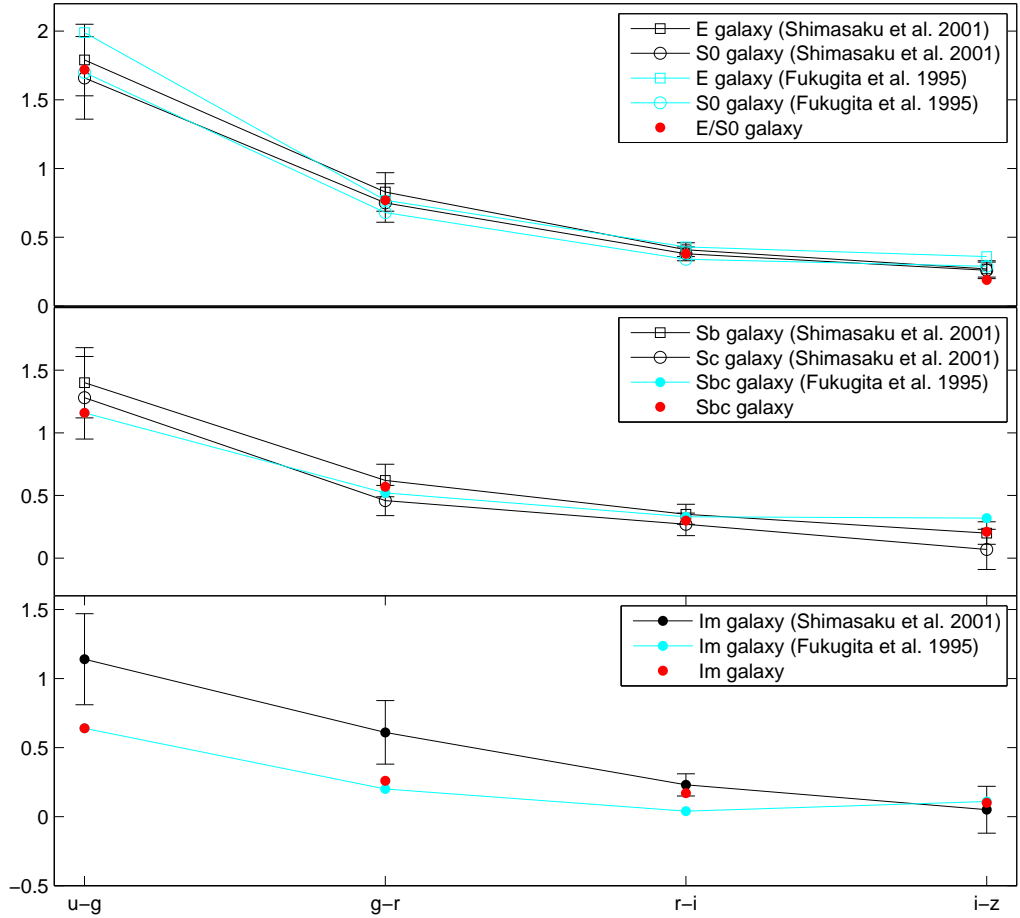}
 \caption{\footnotesize SDSS colors for three galaxies of different
   spectromorphological types. Black symbols are colors obtained from
   Shimasaku et al. (2001) with a rms error bar, blue ones are
   obtained from Fukugita et al. (1995), and red points are the colors
   obtained by applying the transformation equations of section 4.1 to
   the ALHAMBRA photometry of these three galaxy templates (Benítez et
   al. 2004).}
 \end{figure}
\end{center}

\begin{center}
 \begin{figure}[H]
 \centering
 \includegraphics[width=\columnwidth]{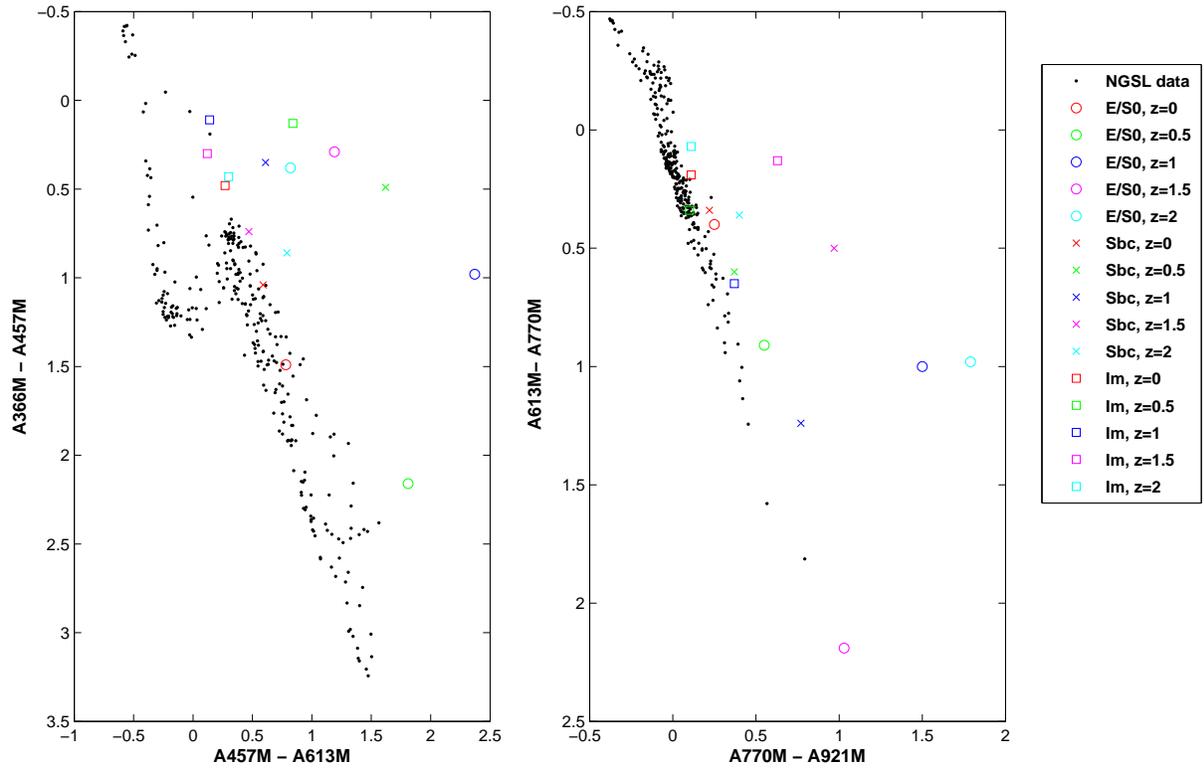}
 \caption{\footnotesize ALHAMBRA color-color diagrams. Black points
   are the set of primary standard stars from the NGSL. Circles
   represent E/S0 galaxy templates, crosses are Sbc galaxy templates
   and squares represent Im galaxy templates, all of them from Benítez
   et al. (2004). Different colors show different values of z.}
 \end{figure}
\end{center}

\begin{center}
 \begin{figure}[H]
 \centering
 \includegraphics[width=\columnwidth]{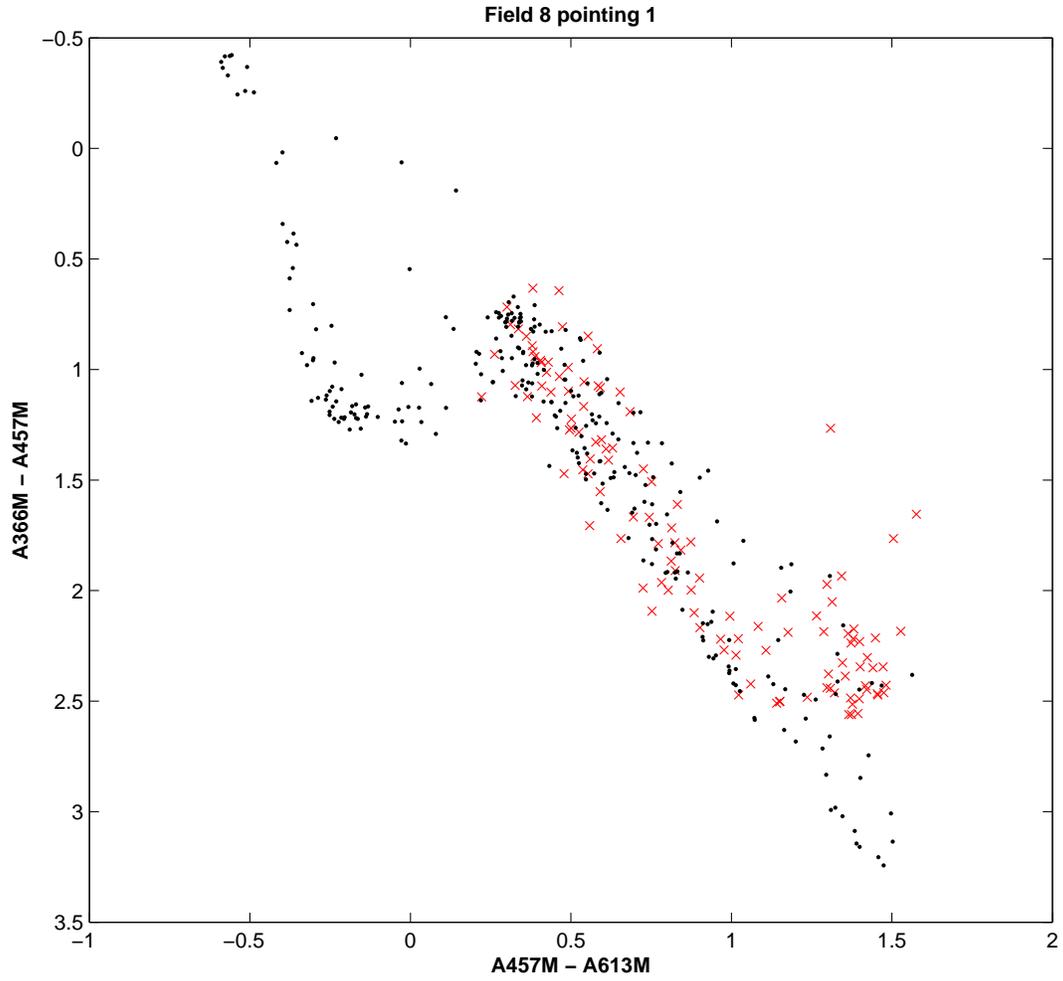}
 \caption{\footnotesize ALHAMBRA color-color diagram. Black points
   represent the 288 stars from the Next Generation Spectral Library,
   and red crosses represent the field stars in field 8 pointing 1 of
   the ALHAMBRA survey.}
 \end{figure}
\end{center}

\begin{center}
 \begin{figure}[H]
  \centering
  \includegraphics[width=\columnwidth]{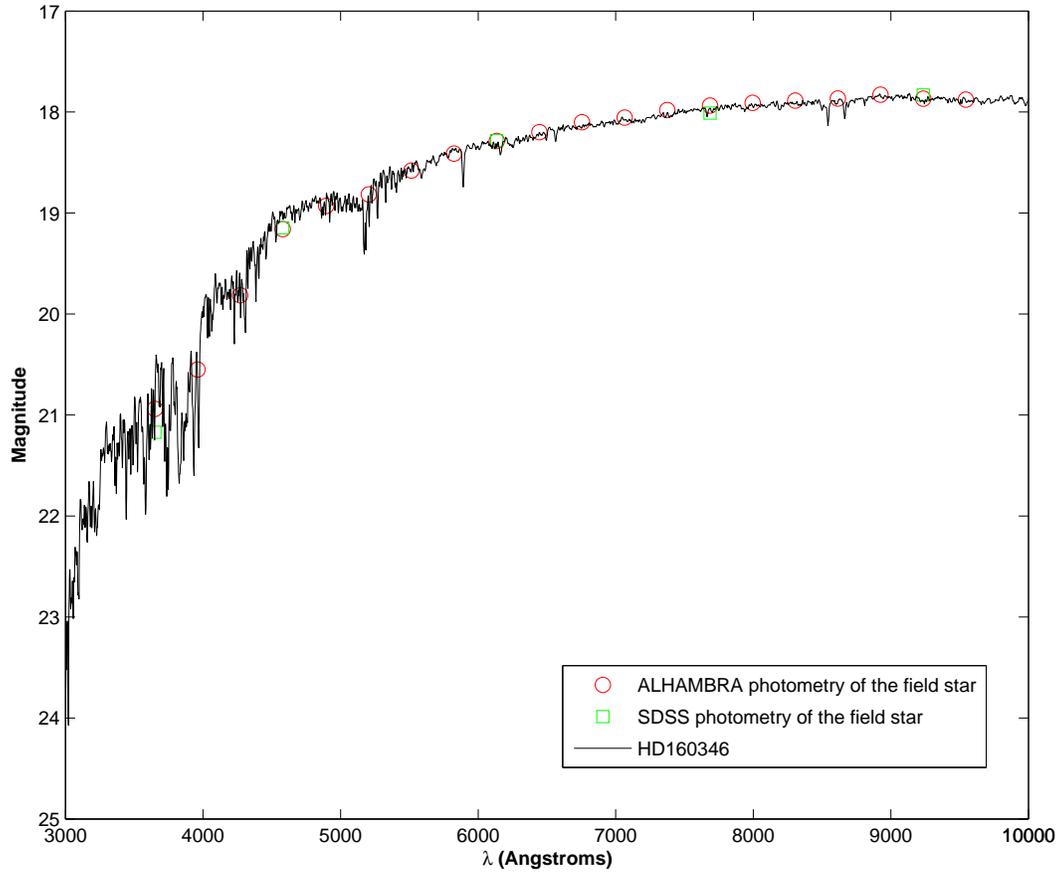}
  \caption{\footnotesize Spectrum in AB magnitudes of the star
    HD160346, belonging to the set of NGSL standard stars. The
    spectrum is scaled so that this object is the best fit for the
    zero point determination to the star at RA = 356.3323 and DEC =
    15.5486. Red circles are the instrumental magnitudes of the field
    star plus the zero points determined with our final strategy, and
    green squares are its SDSS photometry.}
  \end{figure}
 \end{center}

\begin{center}
\begin{table}[H]
\title{\begin{center} Table 13 \end{center}
\begin{center}
 Photometric zero points for ALHAMBRA field 8 pointing 1
\end{center}
}
\begin{center}
 \begin{tabular}{ccc}
\hline \hline
Filter & Zero point & Error \\ \hline
A366M & 7.4229 & 0.0011 \\ 
A394M & 8.3063 & 0.0008 \\ 
A425M & 8.4857 & 0.0007 \\ 
A457M & 8.6686 & 0.0009 \\ 
A491M & 8.7333 & 0.0004 \\ 
A522M & 8.5281 & 0.0005 \\ 
A551M & 8.5879 & 0.0004 \\ 
A581M & 8.6698 & 0.0006 \\ 
A613M & 8.4729 & 0.0004 \\ 
A646M & 8.5864 & 0.0004 \\
A678M & 8.5385 & 0.0005 \\
A708M & 8.3790 & 0.0004 \\
A739M & 8.8394 & 0.0004 \\
A770M & 7.9635 & 0.0003 \\
A802M & 8.4134 & 0.0004 \\
A829M & 8.4139 & 0.0005 \\
A861M & 8.1149 & 0.0004 \\
A892M & 7.2116 & 0.0004 \\
A921M & 6.5880 & 0.0006 \\
A948M & 5.9750 & 0.0006 \\ \hline \hline
\end{tabular}
\end{center}
\end{table}
\end{center}

\begin{center}
 \begin{figure}[H]
   \centering
   \includegraphics[width=\columnwidth]{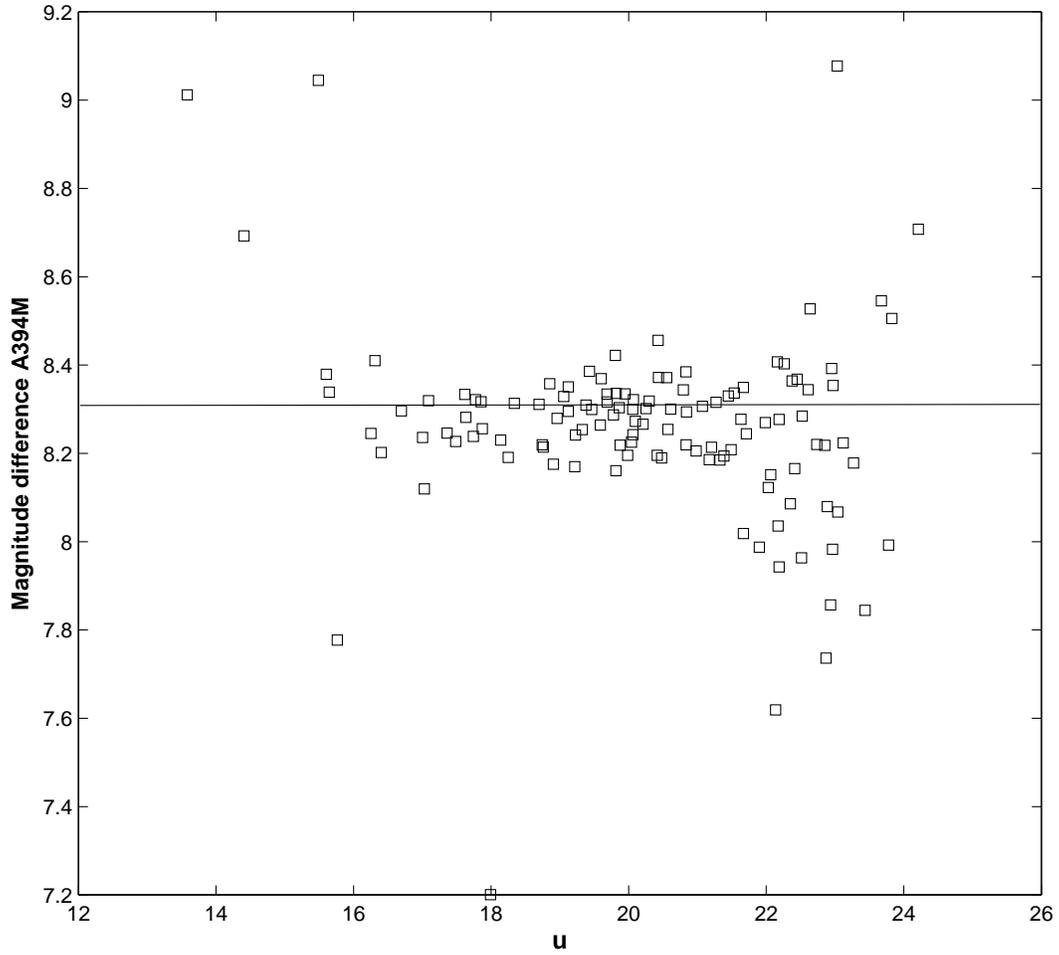}
   \caption{\footnotesize Magnitude differences in filter A394M versus
     SDSS magnitude u, for the field stars in the ALHAMBRA field 8
     pointing 1. They show the difference between the instrumental
     magnitude of the stars and the magnitude obtained from the
     ALHAMBRA-SDSS transformation equations. The horizontal line
     represents the zero-point solution determined with our chosen
     methodology.}
  \end{figure}
\end{center}

\end{document}